\documentclass[a4paper, 11pt]{article}

\usepackage{a4wide}
\usepackage{latexsym}
\usepackage{graphicx}
\usepackage{color}
\usepackage{hyperref}
\usepackage{verbatim}

\title{A stochastic evolutionary model for survival dynamics}

\author{Trevor Fenner, Mark Levene, and George Loizou \\
Department of Computer Science and Information Systems \\
Birkbeck, University of London \\
London WC1E 7HX, U.K. \\ \{trevor,mark,george\}@dcs.bbk.ac.uk}

\date{}

\begin{document}

\maketitle

\begin{abstract}

The recent interest in human dynamics has led researchers to investigate the stochastic processes that explain human behaviour in different contexts. Here we propose a generative model to capture the essential dynamics of survival analysis, traditionally employed in clinical trials and reliability analysis in engineering. In our model, the only implicit assumption made is that the longer an actor has been in the system, the more likely it is to have failed. We derive a power-law distribution for the process and provide preliminary empirical evidence for the validity of the model from two well-known survival analysis data sets.

\end{abstract}

\section{Introduction}

Recent interest in complex systems, such as social networks, the world-wide-web, email networks and mobile phone networks \cite{BARA07},
has led researchers to investigate the processes that may explain the dynamics of human behaviour within these networks.
For example, Barab\'asi \cite{BARA05} suggested that the bursty nature of human behaviour, for example when measuring the inter-event response time of email communication, is a result of a decision-based queuing process. In particular, humans tend to prioritise actions, for example when deciding which email to respond to, and therefore a priority queue model was proposed in \cite{BARA05}, leading to a heavy-tailed power-law distribution of inter-event times.

\smallskip

The tail of a power-law distribution decays polynomially in contrast to the exponential decay characteristic of the Gaussian distribution, which is why it is also referred to as a heavy-tailed distribution. A {\em power-law} distribution takes the general form
\begin{displaymath}
g(i) = \frac{C}{i^{\rho}},
\end{displaymath}
representing the proportion of observations having the value $i$, where $C$ and $\rho$ are positive constants; we call $\rho$ the {\em exponent} of the distribution \cite{NEWM05}.

\smallskip

Survival analysis \cite{KLEI11} provides statistical methods to estimate the time until an event will occur, known as the {\em survival time}. Typically, an event in a survival model is referred to as a {\em failure}, as it often has negative connotations, such as death or the contraction of a disease, although it could also be positive, such as the time to return to work or to recover from a disease. In the context of email communication mentioned above, an event might be a reply  to an email. Traditional applications of survival analysis are in clinical trials \cite{FLEM00}, and reliability engineering \cite{MA08}, the analogue of survival analysis for mechanical systems. However, one can envisage that survival analysis would find application in newer human dynamics scenarios in complex systems, such as those arising in social and communication networks \cite{BARA05,CAND08,ZHOU08}.

\smallskip

Of particular interest to us has been the formulation of a {\em generative model} in the form of a stochastic process by which a complex system evolves and gives rise to a power law or other distribution \cite{FENN02,FENN07,FENN12}.
This type of research builds on the early work of Simon \cite{SIMO55}, and the more recent work of Barab\'asi's group \cite{ALBE01} and other researchers \cite{BORN07a}. In the context of human dynamics, the priority queue model \cite{BARA05} mentioned above is a generative model characterised by a heavy-tailed distribution. In the bigger picture, one can view the goal of such research as being similar to that of {\em social mechanisms} \cite{HEDS98}, which looks into the processes, or mechanisms, that can explain observed social phenomena.
Using an example given in \cite{SCHE98a}, the growth in the sales of a book can be explained by the well-known logistic growth model \cite{TSOU02}.

\smallskip

The motivation of this paper is in the formulation of a simple generative model that will capture the essential dynamics of survival analysis applications. For this purpose, we make use of an urn-based stochastic model, where the actors are called {\em balls}, and a ball being present in $urn_i$, the $i$th urn, indicates that the actor represented by the ball has so far survived for $i$ time steps. An actor could, for example, be a subject in a clinical trial or an email that has not yet been replied to. As a simplification, we assume that time is discrete and that, at any given time, one ball may join the system with a fixed probability.  As a result, at any given time, say $t$, we may have at most one ball in $urn_i$, for all $i \le t$. Alternatively, with a fixed probability, an existing ball in the system may be chosen uniformly at random and discarded. We note that at any time $t$, if $i < j$, the probability that $urn_i$ is empty is less than the probability that $urn_j$ is empty,
as a ball in $urn_j$ could have been discarded at any of the previous $j$ time steps, whereas a ball in $urn_i$ could have only been discarded at any of the previous $i$ time steps, and a ball to be discarded is chosen uniformly.

\smallskip

This mechanism can be contrasted with the preferential attachment rule in evolving networks \cite{ALBE01} (also known as the ``rich get richer'' phenomenon \cite{BARA07}), where the probability of adding a new link to a node (or actor) is proportional
to the number of existing links that the node already has. In our case, actors are chosen uniformly rather than preferentially; however, the model keeps a record of the time for which an actor has survived so for.

\smallskip

Our main result is to derive a power-law distribution for the probability that, after $t$ steps of the stochastic process outlined above, there is a surviving ball in $urn_i$, where $i \le t$. Thus, in our model, the {\em survivor function} \cite{KLEI11}, which gives the probability that a patient (in our model a ball) survives for more than a given time, can be approximated by a power-law distribution.
It is interesting to observe that the resulting distribution has two parameters, $i$ and $t$, as in \cite{FENN12}, whereas
most previously studied generative stochastic models \cite{ALBE01,NEWM05}, including those in our previous work \cite{FENN02,FENN07}, result in steady state distributions that are asymptotic in $t$ to a distribution with a single parameter $i$.
As a proof of concept, we demonstrate the validity of our model by analysing two well-known data sets from the survival analysis literature \cite{KLEI11}.

\smallskip

The rest of the paper is organised as follows. In Section~\ref{sec:urn}, we present our stochastic urn-based model that provides us with a mechanism to model the essential dynamics of survival models, and we derive the resulting power-law distribution. In Section~\ref{sec:survival}, we apply our generative model to two well-known data sets from survival analysis, and finally, in Section~\ref{sec:conc}, we give our concluding remarks.

\section{An Evolutionary Urn Transfer Model}
\label{sec:urn}

In this section we formalise our stochastic urn model for modelling the dynamic aspects of a time-varying system and present an approximate solution to the mean field equations describing the model.

\smallskip

We assume a countable number of urns, $urn_1, urn_2, \ldots \ $. Initially all the urns are empty except $urn_1$, which has one ball in it. Let $F_i(t)$ be 1 or 0, respectively, according to whether or not there is a ball in $urn_i$ at time $t$ of the stochastic process. Initially we set $F_1(1) = 1$, and for all other urns $F_i(1) = 0$. The {\em age} of a ball in $urn_i$ is defined to be $i$. Then, at time $t+1$ of the stochastic process, where $t \ge 1$, one of two things may occur:

\renewcommand{\labelenumi}{(\roman{enumi})}
\begin{enumerate}
\item with probability $p$, where $0 < p < 1$, a new ball is put into $urn_1$ (i.e its initial age is 1), or

\item with probability $q (1 - p)$, where $0 < q < 1$, an urn is selected, with $urn_i$ being selected with probability proportional to $F_i(t)$, i.e. one of the non-empty urns is selected uniformly at random, and the ball in the selected urn is discarded.
\end{enumerate}

Next, the age of all balls remaining in the system, apart from a new ball that may have just been put into $urn_1$ during this time step, is incremented by 1, i.e. any ball in $urn_i$ is moved to $urn_{i+1}$ for each $i$.

\medskip

We observe that, in this model, $F_i(t)$ is equal to either 1 or 0, since at most one new ball is generated at time $t$ and the age of all other balls increases by one at time $t$. Moreover, it can be seen that, at any given time $t > 1$, the probability of there being a ball in $urn_1$ is $p$.

\smallskip

We constrain the system so that on average more balls are added to the system than are discarded, i.e
\begin{displaymath}
p > q (1-p),
\end{displaymath}
otherwise the system would almost surely degenerate into a state of emptiness, i.e. having no balls in the system. This constraint is an instance of the gambler's ruin problem \cite{ROSS96}, from which it follows that the probability that the urn transfer process will {\em not} terminate with all the urns being empty is strictly positive \cite{FENN02}.

\medskip

Now let $f_i(t) = E(F_i(t))$. As $F_i(t)$ is either $0$ or $1$, this expectation is equal to the probability that there is a ball in $urn_i$ at time $t$; so $f_1(t) = p$.
So, the expected total age of the balls in the system for large $t$ is
\begin{displaymath}
E\left(\sum_{i=1}^t i F_i(t)\right) = \sum_{i=1}^t i f_i(t).
\end{displaymath}

Let $\tau = p - q(1-p)$, then $\tau t$ is the expected number of balls in the system at time $t$.
For sufficiently large $t$, we obtain the following simple bounds on the expected total age of the balls,
\begin{equation}\label{eq:ifi}
\frac{\tau^2 t^2}{2} \le \sum_{i=1}^t i f_i(t) \le  \frac{(2 - \tau) \tau t^2}{2},
\end{equation}
so it follows, on dividing by $\tau t$, that the average age of a ball is proportional to $t$.
To obtain the lower bound in (\ref{eq:ifi}), we assume that the $\tau t$ balls in the system are as young as possible;
conversely, to obtain the upper bound, we assume that they are as old as possible.

\smallskip

Following \cite{FENN02}, we now state the mean field equations for the urn transfer model. At time $t > 1$, for $i > 1$ we have
\begin{equation}\label{eq:a1}
f_{i}(t) = f_{i-1}(t-1) -  q (1-p) \beta_{t-1} f_{i-1}(t-1),
\end{equation}
where $\beta_t$ is the probability of choosing any particular non-empty urn in step (ii) at time $t$; this is given by
\begin{equation}\label{eq:betat}
\beta_t = E \left (\frac{1}{\sum_{j=1}^t F_j(t)} \right) \approx \frac{1}{\sum_{j=1}^t f_j(t)} = \frac{1}{\tau t},
\end{equation}
since $\tau t$ is the expected number of balls in the system at time $t$.

\smallskip

In the boundary case, when $i = 1$, we have
\begin{equation}\label{eq:p}
f_1(t) = p.
\end{equation}
\smallskip

We can now rewrite (\ref{eq:a1}) as
\begin{equation}\label{eq:a3}
f_{i}(t) = \left( 1 -  \frac{\kappa}{t-1} \right) f_{i-1}(t-1),
\end{equation}
where
\begin{equation}\label{eq:kappa}
\kappa = \frac{q(1-p)}{\tau} = \frac{q(1-p)}{p - q(1-p)} = \frac{p - \tau}{\tau}.
\end{equation}
\smallskip

Note that, when $i > t$, we have $f_i(t) = 0$.
We can solve equations (\ref{eq:p}) and (\ref{eq:a3}) directly, obtaining
\begin{equation}\label{eq:a4}
f_{i}(t) = p
\left( 1 - \frac{\kappa}{t-1} \right)
\left( 1 - \frac{\kappa}{t-2} \right) \cdots
\left( 1 - \frac{\kappa}{t-(i-2)} \right)
\left( 1 - \frac{\kappa}{t-(i-1)} \right).
\end{equation}

We first rewrite (\ref{eq:a4}) as
\begin{equation}\label{eq:a5}
f_{i}(t) = p
\left( \frac{t-1-\kappa}{t-1} \right)
\left( \frac{t-2-\kappa}{t-2} \right) \cdots
\left( \frac{t-i+2-\kappa}{t-i+2} \right)
\left( \frac{t-i+1-\kappa}{t-i+1} \right).
\end{equation}

Finally, rewriting (\ref{eq:a4}) in terms of the Gamma function $\Gamma(\cdot)$ \cite{GRAH94}, we obtain
\begin{displaymath}
f_{i}(t) =
\frac{p \ \Gamma(t - \kappa) \ \Gamma(t-i+1)}{\Gamma(t) \ \Gamma(t-i+1 - \kappa)}.
\end{displaymath}
\smallskip

Making use of the limit (see \cite[6.1.46]{ABRA72}),
\begin{displaymath}
\lim_{x\to\infty} \left[  x^{b-a} \frac{\Gamma(x+a)}{\Gamma(x+b)} \right] = 1,
\end{displaymath}
$f_{i}(t)$ can be approximated by
\begin{equation}\label{eq:a7}
f_{i}(t) \approx p \ t^{-\kappa} \left( t-i \right)^{\kappa} = p \left( 1 - \frac{i}{t} \right)^{\kappa} = p \left( 1 - \frac{i}{t} \right)^{\frac{q (1-p)}{p-q(1-p)}},
\end{equation}
provided $t-i$ is ``large'', more specifically that $t-i \to\infty$ as $t \to\infty$.

\smallskip

Now let $\omega_t$ be a slowly increasing function of $t$, i.e. $\omega_t \to\infty$ as $t \to\infty$, for example,
$\omega_t = \ln t$. Then (\ref{eq:a7}) holds provided $i \le t - \omega_t$. Since, from (\ref{eq:a5}), it follows that $f_i(t)$ is strictly monotonically decreasing in $i$, when $i > t - \omega_t$, we have
\begin{displaymath}
f_i(t) < f_{t-\omega_t}(t) \approx p \left(\frac{\omega_t}{t}\right)^\kappa.
\end{displaymath}

It follows that the expected number of balls $T$ in urns for which (\ref{eq:a7}) may not hold, i.e. in the tail defined by $t - \omega_t < i \le t$, is bounded above by
\begin{equation}\label{eq:a8}
T = \sum_{i=t-\omega_t+1}^t f_i(t) < \frac{p \ {\omega_t}^{\kappa +1}}{t^\kappa}.
\end{equation}
\smallskip

By approximating the sums by integrals, it is not difficult to show that the approximation to $f_i(t)$ in (\ref{eq:a7}) satisfies
\begin{displaymath}
\sum_{i=1}^t   p \left( 1 - \frac{i}{t} \right)^{\kappa} \approx \tau t
\qquad{\rm{and}}\qquad \sum_{i=1}^t i   p \left( 1 - \frac{i}{t} \right)^{\kappa} \approx \frac{\tau^2 t^2}{p+\tau}.
\end{displaymath}

These are consistent with the expected number of balls being $\tau t$
and the bounds on the expected total age of the balls in (\ref{eq:ifi}).

\medskip

In survival analysis, briefly described in the next section, we are often interested in the {\em survivor function} $S(\theta)$ \cite{KLEI11}, which gives the probability that a patient in a given study survives for longer than a specified time $\theta$. The survivor function is usually estimated via a step function by computing the probability that a patient survives until time $\theta$, for $\theta = 1,2,\ldots, t$. This step function is known as the {\em Kaplan-Meier estimator} \cite{KAPL58,KLEI11}.
By comparing (\ref{eq:a4}) and the Kaplan-Meier estimate for the survivor function \cite[equation~(2b)]{KAPL58}, this estimate is seen to be analogous to $f_i(t)$ for an actor that was born at time $t-i$; more specifically, we can approximate $S(i)$ by $f_i(t)/p$.

\smallskip

We note that although in theory the survivor function $S(\theta)$ does not depend on the length $t$ of the trial, in practice the Kaplan-Meier estimate will be more accurate for longer trials. However, the Kaplan-Meier estimate is most accurate when most of the patients are still present in the study, since, when there are only a few patients left, the estimate may be inaccurate \cite{RICH10}. This is consistent with our approximation of $f_i(t)$, which holds when $t-i$ is large.

\smallskip

The Kaplan-Meier estimator also takes into account {\em censored} data \cite{KLEI11}, when, for example, a patient drops out before the end of the study period. Although our evolutionary urn transfer model does not include censoring, it could be incorporated by allowing the possibility that when a ball is removed from an urn it is not necessarily counted as being discarded. We further note that, while in traditional survivor models patients join a study in batches, in our model individual balls continue to join the system with a fixed probability. Our model could be generalised to allow several balls to join the system at any given time, and also by letting the arrival probability $p$ depend on $t$; we leave consideration of such enhancements for future work.

\section{Application to Survival Analysis}
\label{sec:survival}

Survival analysis is well established within the statistics community, dealing with the analysis of {\em time-to-event} data \cite{KLEI11}.
For example, in clinical trials, one is monitoring patients and how likely they are to survive or to respond to a treatment within a given time frame.

\smallskip

In our model, the objects being monitored are represented by balls and they are considered to have survived for as long as they remain in the system. The death event is modelled by discarding a ball, and the arrival event is modelled by putting a new ball into the first urn.

\smallskip

Our stochastic model has three input parameters: the arrival rate $p$, the death rate $q(1-p)$, and the time $t$ at which the system is observed. Given these parameters, the survival probability of a ball in $urn_i$ at time $t$, where $i \le t$, is approximated by $f_i(t)$ as given by (\ref{eq:a7}). In other words, given $t$, $f_i(t)$ is the probability that a ball that enters the system at time $t-i$ survives for at least $i$ steps before it is discarded.

\smallskip

As a proof of concept, we make use of two well-known data sets from survival analysis: (i) the {\em Vets} data set \cite[p.~72]{KLEI11}, which contains the survival times in days for 137 lung cancer patients from the Veteran’s Administration Lung Cancer Trial, and (ii) the {\em Addicts} data set \cite[p.~91]{KLEI11}, which contains the times in days spent by 238 heroin addicts in one of two methadone clinics.
The longest non-censored actor survived for 999 days for the Vets data and for 899 days for the Addicts data. Moreover, the last censored actor
was at 231 days for the Vets data and at 1076 days for the Addicts data.

\smallskip

We first computed the Kaplan-Meier curves from the raw Vets and Addicts data sets. Then, recalling that we approximate the survivor function $S(i)$ by $f_i(t)/p$, using Matlab, we derived $p$ and $q$ using non-linear least squares regression on the Kaplan-Meier curves, assuming the distribution for $f_i(t)$ given in (\ref{eq:a7}).
To obtain consistency for $f_i(t)$, the approximate survivor function $S(i)$ was multiplied by an estimate of $p$, which was computed as the ratio of the number of actors to the number of days of the trial; for the Vets data set this estimate was $137/999 = 0.1371$ and for the Addicts data it was $238/1076 = 0.2212$. The values for $p$, $q$ and $\kappa$ that we obtained, together with the coefficient of determination $R^2$ \cite{MOTU95}, are shown in the rows of Table~\ref{table:kappa} captioned {\em Vets data} and {\em Addicts data}.

\smallskip

To test the validity of the model, we then carried out simulations in Matlab of the stochastic urn transfer model using the values of $p$ and $q$ obtained from the above regression. The simulations for the Vets data set was run for 1000 steps and for the Addicts data for 900 steps, and these were repeated one million times. We then calculated the average value of $f_i(t)$ for $i=1,2,\ldots,t$, over the one million runs.

\smallskip

\begin{figure}[ht]
\centerline{\includegraphics[width=15cm,height=10cm]{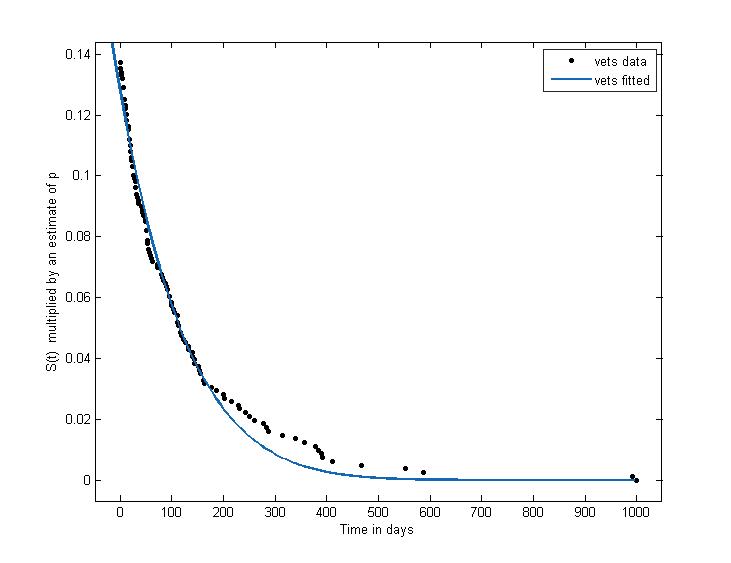}}
\caption{\label{fig:vets} The fitted curve for the Kaplan-Meier curve of the Vets data set.}
\end{figure}

\begin{figure}[ht]
\centerline{\includegraphics[width=15cm,height=10cm]{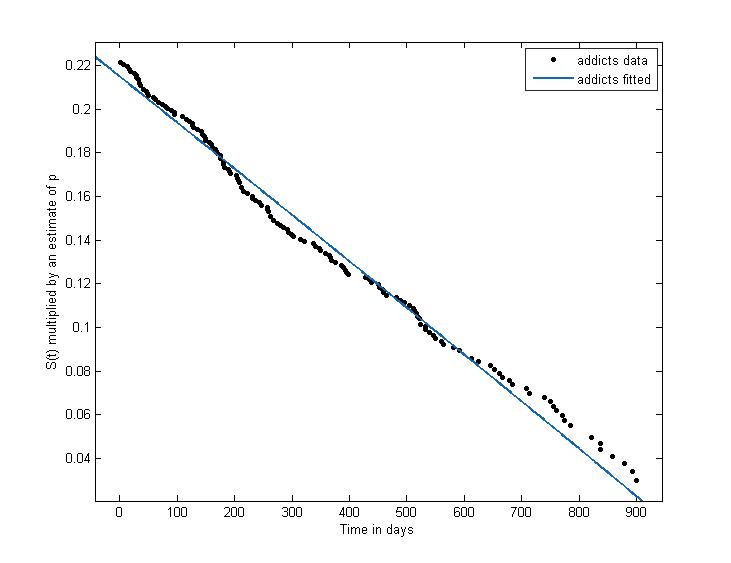}}
\caption{\label{fig:addicts} The fitted curve for the Kaplan-Meier curve of the Addicts data set.}
\end{figure}

We obtained new values for $p$ and $\kappa$ using non-linear regression on the resulting averaged data set, assuming the distribution in (\ref{eq:a7}). We then computed $q$ from $p$ and $\kappa$ using (\ref{eq:kappa}). These values are shown in the rows of Table~\ref{table:kappa} captioned {\em Vets sim} and {\em Addicts sim}, together with the $R^2$ values. It can be seen that the values of $p$ and $q$ from the simulation closely match their counterparts obtained directly from the raw data.
The $\kappa$ values are also close, although we note that $\kappa$ is quite sensitive to small changes in the values of $p$ and $q$.
We also show the expected number of balls in the tail $T$, given by (\ref{eq:a8}), where $\omega_t = \ln t$. The tail for the Addicts data is much larger than that for the Vets data, although, even in this latter case, the expected fraction of balls in the tail is less than $0.01\%$. The Kaplan-Meier curve from the raw data and the fitted regression curve from the model are shown in Figure \ref{fig:vets} for the Vets data and in Figure \ref{fig:addicts} for the Addicts data.

\begin{table}[ht]
\begin{center}
\begin{tabular}{|c|c|c|c|c|c|c|}\hline
Data set     & $p$    & $q$    & $\kappa$ & $R^2$  & $t$  & {\em T}  \\ \hline \hline
Vets data    & 0.1277 & 0.1294 & 7.6132   & 0.9862 & 1000 & 7.089400e-18 \\ \hline
Vets sim     & 0.1230 & 0.1240 & 7.6094   & 0.9985 & 1000 & 6.960000e-18 \\ \hline \hline
Addicts data & 0.2148 & 0.1355 & 0.9813   & 0.9885 & 900  & 7.384400e-03 \\ \hline
Addicts sim  & 0.2141 & 0.1351 & 0.9836   & 0.9995 & 900  & 7.271500e-03 \\ \hline
\end{tabular}
\end{center}
\caption{\label{table:kappa} Values for $p$ and $q$ fitted from the data and recovered from the simulations.}
\end{table}
\smallskip

\smallskip

\smallskip

\section{Concluding Remarks}
\label{sec:conc}

We have proposed a stochastic evolutionary urn model for survival analysis applications in the context of human dynamics.
In our model, actors (represented by balls) remain in the system and survive until they die (i.e. are discarded).
At any given time step each actor currently in the system is equally likely to be discarded,
although in practice there may be many other factors that influence the survival of an actor.
Equation~(\ref{eq:a7}) exhibits the power law in $(t-i)/t$, the proportion of the total time period before the actor joined the system. Despite the simplicity of our model, the results we have presented provide preliminary evidence for its validity, although further validation and possible enhancements to the model, as indicated at the end of Section~\ref{sec:urn}, would be desirable. Generative models, such as the one we have presented, have the potential to explain observed social phenomena and, in this context, social mechanisms, as discussed in the introduction. Moreover, they allow us to gain insight into the underlying processes and may also be useful for prediction purposes \cite{HEND01}.


\bibliographystyle{alpha}

\begin{thebibliography}{CGW{\etalchar{+}}08}

\bibitem[AB02]{ALBE01}
R.~Albert and A.-L. Barab\'asi.
\newblock Statistical mechanics of complex networks.
\newblock {\em Reviews of Modern Physics}, 74:47--97, 2002.

\bibitem[AS72]{ABRA72}
M.~Abramowitz and I.A. Stegun, editors.
\newblock {\em Handbook of Mathematical Functions with Formulas, Graphs and
  Mathematical Tables}.
\newblock Dover, New York, NY, 1972.

\bibitem[Bar05]{BARA05}
A.-L. Barab\'asi.
\newblock The origin of bursts and heavy tails in human dynamics.
\newblock {\em Nature}, 435:207--211, 2005.

\bibitem[Bar07]{BARA07}
A.-L. Barab\'asi.
\newblock The architecture of complexity: {F}rom network strucutre to human
  dynamics.
\newblock {\em IEEE Control Systems Magazine}, 27:33--42, 2007.

\bibitem[BSV07]{BORN07a}
S.~B\"{o}rner, S.~Sanyal, and A.~Vespignani.
\newblock Network science.
\newblock {\em Annual Review of Information Science \& Technology (ARIST)},
  41:537--607, 2007.

\bibitem[CGW{\etalchar{+}}08]{CAND08}
J.~Candia, M.C. Gonz\'alez, P.~Wang, T.~Schoenhar, G.~Madey, and A.-L.
  Barab\'asi.
\newblock Uncovering individual and collective human dynamics from mobile phone
  records.
\newblock {\em Journal of Physics A: Mathemtical and Theoretical}, 41:224015,
  11pp, 2008.

\bibitem[FL00]{FLEM00}
T.R. Fleming and D.Y. Lin.
\newblock Survival analysis in clinical trials: {P}ast developments and future
  directions.
\newblock {\em Biometrics}, 56:971--983, 2000.

\bibitem[FLL05]{FENN02}
T.I. Fenner, M.~Levene, and G.~Loizou.
\newblock A stochastic evolutionary model exhibiting power-law behaviour with
  an exponential cutoff.
\newblock {\em Physica A}, 335:641--656, 2005.

\bibitem[FLL12]{FENN12}
T.I. Fenner, M.~Levene, and G.~Loizou.
\newblock A discrete evolutionary model for chess players’ ratings.
\newblock {\em IEEE Transactions on Computational Intelligence and AI in
  Games}, 4:84--93, 2012.

\bibitem[FLLR07]{FENN07}
T.I. Fenner, M.~Levene, G.~Loizou, and G.~Roussos.
\newblock A stochastic evolutionary growth model for social networks.
\newblock {\em Computer Networks}, 51:4586--4595, 2007.

\bibitem[GKP94]{GRAH94}
R.L. Graham, D.E. Knuth, and O.~Patachnik.
\newblock {\em Concrete Mathematics: A Foundation for Computer Science}.
\newblock Addison-Wesley, Reading, Ma., 2nd edition, 1994.

\bibitem[HJS01]{HEND01}
R.~Henderson, M.~Jones, and J.~Stare.
\newblock Accuracy of point predictions in survival analysis.
\newblock {\em Statistics in Medicine}, 20:3083--3096, 2001.

\bibitem[HS98]{HEDS98}
P.~Hedstr\"{o}m and R.~Swedberg.
\newblock Social mechanisms: {A}n introductory essay.
\newblock In P.~Hedstr\"{o}m and R.~Swedberg, editors, {\em Social Mechanisms:
  An Analytical Approach to Social Theory}, pages 1--31. Cambridge University
  Press, Cambridge, U.K., 1998.

\bibitem[KK11]{KLEI11}
D.G. Kleinbaum and M.~Klein.
\newblock {\em Survival Analysis, A Self-Learning Text}.
\newblock Springer Science+Business Media, LLC, New York, NY, 3rd edition,
  2011.

\bibitem[KM58]{KAPL58}
E.L. Kaplan and P.~Meier.
\newblock Nonparametric estimation from incomplete observations.
\newblock {\em Journal of the American Statistical Association}, 53:457--481,
  1958.

\bibitem[MK08]{MA08}
Z.~Ma and A.W. Krings.
\newblock Survival analysis approach to reliability, survivability and
  prognostics and health management ({PHM}).
\newblock In {\em Proceedings of the IEEE Aerospace Conference}, pages 1--20,
  Big Sky, MT, 2008.

\bibitem[Mot95]{MOTU95}
H.~Motulsky.
\newblock {\em Intuitive Biostatistics}.
\newblock Oxford University Press, Oxford, 1995.

\bibitem[New05]{NEWM05}
M.E.J. Newman.
\newblock Power laws, {P}areto distributions and {Z}ipf's law.
\newblock {\em Contemporary Physics}, 46:323--351, 2005.

\bibitem[RNP{\etalchar{+}}10]{RICH10}
J.T. Rich, J.G. Neely, R.C. Paniello, C.C.J. Voelker, B.~Nussenbaum, and E.W.
  Wang.
\newblock A practical guide to understanding {Kaplan-Meier} curves.
\newblock {\em Otolaryngology–Head and Neck Surgery}, 143:331--336, 2010.

\bibitem[Ros96]{ROSS96}
S.M. Ross.
\newblock {\em Stochastic Processes}.
\newblock John Wiley {\&} Sons, New York, NY, 2nd edition, 1996.

\bibitem[Sch98]{SCHE98a}
T.C. Schelling.
\newblock Social mechanisms and social dynamics.
\newblock In P.~Hedstr\"{o}m and R.~Swedberg, editors, {\em Social Mechanisms:
  An Analytical Approach to Social Theory}, pages 32--44. Cambridge University
  Press, Cambridge, U.K., 1998.

\bibitem[Sim55]{SIMO55}
H.A. Simon.
\newblock On a class of skew distribution functions.
\newblock {\em Biometrika}, 42:425--440, 1955.

\bibitem[TW02]{TSOU02}
A.~Tsoularis and J.~Wallace.
\newblock Analysis of logistic growth models.
\newblock {\em Mathematical Biosciences}, 179:21--55, 2002.

\bibitem[ZHW08]{ZHOU08}
T.~Zhou, {X.-P.} Han, and {B.-H.} Wang.
\newblock Towards the understanding of human dynamics.
\newblock In M.~Burguete and L.~Lam, editors, {\em Science Matters},
  chapter~12, pages 207--233. World Scientific, Singapore, 2008.

\end{thebibliography}
\newcommand{\etalchar}[1]{$^{#1}$}

\end{document}